\documentclass[aip,apl,reprint,superscriptaddress,twocolumn]{revtex4-1}

\usepackage[none]{hyphenat}
\usepackage{graphicx}
\usepackage{dcolumn}
\usepackage{bm}
\usepackage{amsmath}
\usepackage{color}
\usepackage[colorlinks,
linkcolor=blue,
citecolor=blue,
urlcolor=blue]{hyperref}
\makeatother

\begin{document}
\title{Characteristics of Branched Flows of High-Current Relativistic Electron Beams in Porous Materials}

\author{K. Jiang}
\affiliation{Shenzhen Key Laboratory of Ultraintense Laser and Advanced Material Technology, Center for Advanced Material Diagnostic Technology, and College of Engineering Physics, Shenzhen Technology University, Shenzhen 518118, People's Republic of China}

\author{T. W. Huang}
\email{taiwu.huang@sztu.edu.cn}
\affiliation{Shenzhen Key Laboratory of Ultraintense Laser and Advanced Material Technology, Center for Advanced Material Diagnostic Technology, and College of Engineering Physics, Shenzhen Technology University, Shenzhen 518118, People's Republic of China}

\author{R. Li}
\affiliation{Shenzhen Key Laboratory of Ultraintense Laser and Advanced Material Technology, Center for Advanced Material Diagnostic Technology, and College of Engineering Physics, Shenzhen Technology University, Shenzhen 518118, People's Republic of China}

\author{C. T. Zhou}
\email{zcangtao@sztu.edu.cn}
\affiliation{Shenzhen Key Laboratory of Ultraintense Laser and Advanced Material Technology, Center for Advanced Material Diagnostic Technology, and College of Engineering Physics, Shenzhen Technology University, Shenzhen 518118, People's Republic of China}

\date{\today}
\begin{abstract}
Branched flow is a universal phenomenon in which treebranch-like filaments form through traveling waves or particle flows in irregular mediums. Branched flow of high-current relativistic electron beams (REBs) has been recently discovered [Phys. Rev. Lett. \textbf{130}, 185001 (2023)]. It exhibits unique features, including remarkably high beam density at predictable caustic locations, efficient energy coupling between the beam and background medium, etc. This paper presents investigations on REB branching, focusing on the influence of interaction parameters on branching patterns and providing detailed analyses of the dynamics of individual beam electrons.  The insights gained contribute to a nuanced understanding of the intricate nature of REB branching and its potential applications in the future.

\end{abstract}
\maketitle
\sloppy{}

\section{Introduction}

Branched flow refers to a phenomenon wherein treebranch-like filaments emerge as waves or particle flows propagate through an irregular medium \cite{Heller}. The medium's irregularity manifests through an effective potential $V$, which is randomly uneven, spatially smooth, weak ($V_{\mathrm{rms}}\ll E_{k}$, where $E_{k}$ represents the kinetic energy of the traveling flow), and long-range correlated (with a correlation length $l_{c}$ exceeding the flow wavelength) \cite{Kaplan,Metzger,Metzger1}. In such conditions, sudden and significant momentum changes in the flow are absent, but accumulated small changes profoundly influence the morphologies and dynamics of the flow. In particular, caustics and filaments form as the flow undergoes bending and bundling at favorable locations. For example, instead of smoothly spreading, tsunami waves refract and form branching strands due to fluctuations in the ocean height \cite{Deguelder}. Impurities within semiconductors can induce branched flows of current-carrying electrons \cite{Topinka}. Cosmic rays form branches when traversing inhomogeneous interstellar dust clouds \cite{Pidwerbetsky}. Light undergoes branched flow in complex media characterized by refractive index fluctuations \cite{Patsyk,Patsyk1,Jiang}.

\begin{figure}[!htb] 
	\centering 
	\includegraphics[width=8.6cm]{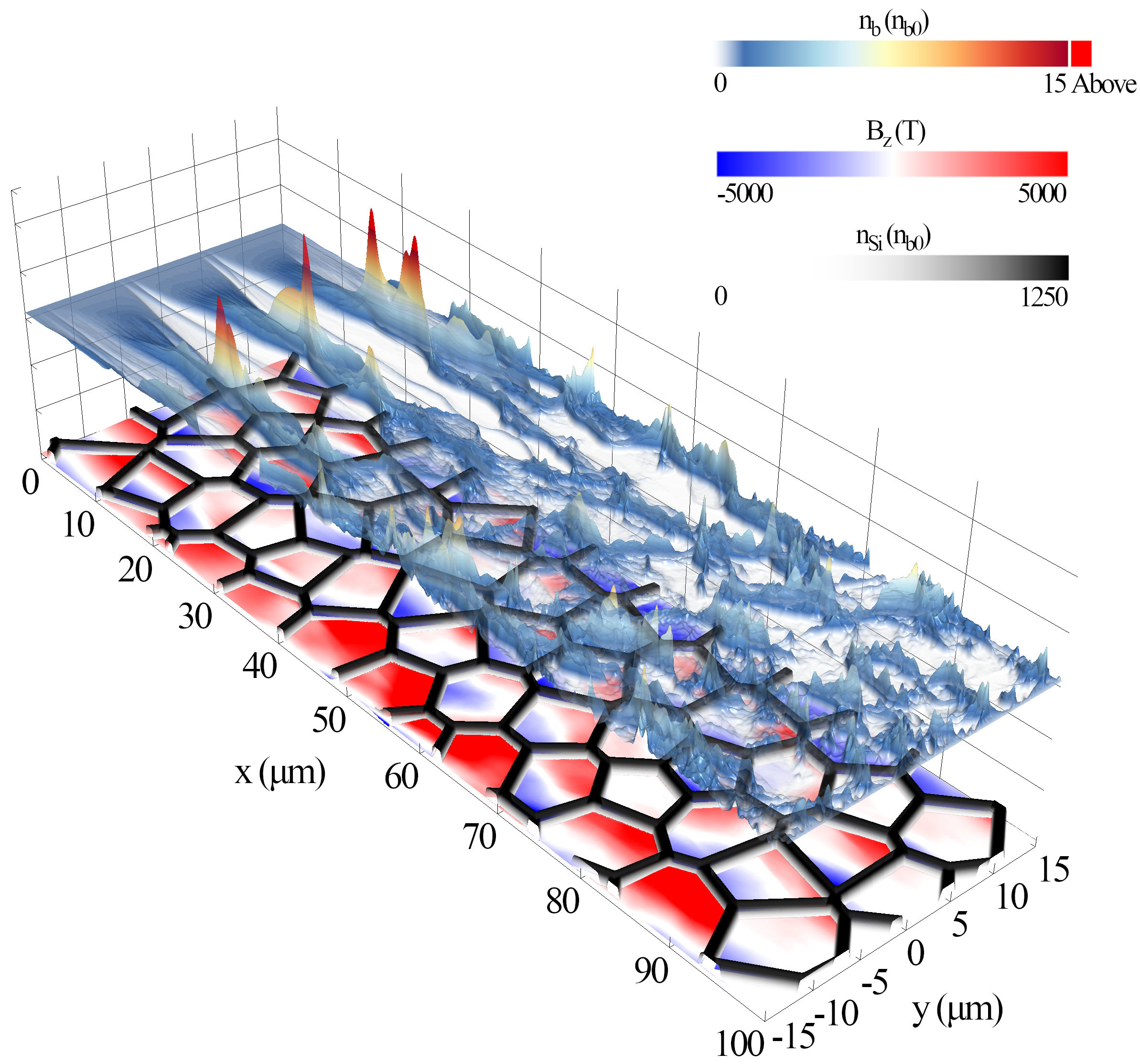} 
	\caption{Formation of branched flow patterns as a high-current relativistic electron beam propagates through a porous SiO$_{2}$ foam along the $x$ direction. The red to blue color bar is for the beam density (in units of $n_{b0}$, same below) at $t=333\ \mathrm{fs}$. The initial density distribution of Si atoms in the foam (black) and self-generated azimuthal magnetic field $B_{z}$ (red and blue, in tesla) at $t=333\ \mathrm{fs}$ are shown beneath the beam density.}
	\label{fig_1}
\end{figure}

\begin{figure*}[!htb]
	\centering 
	\includegraphics[width=14cm]{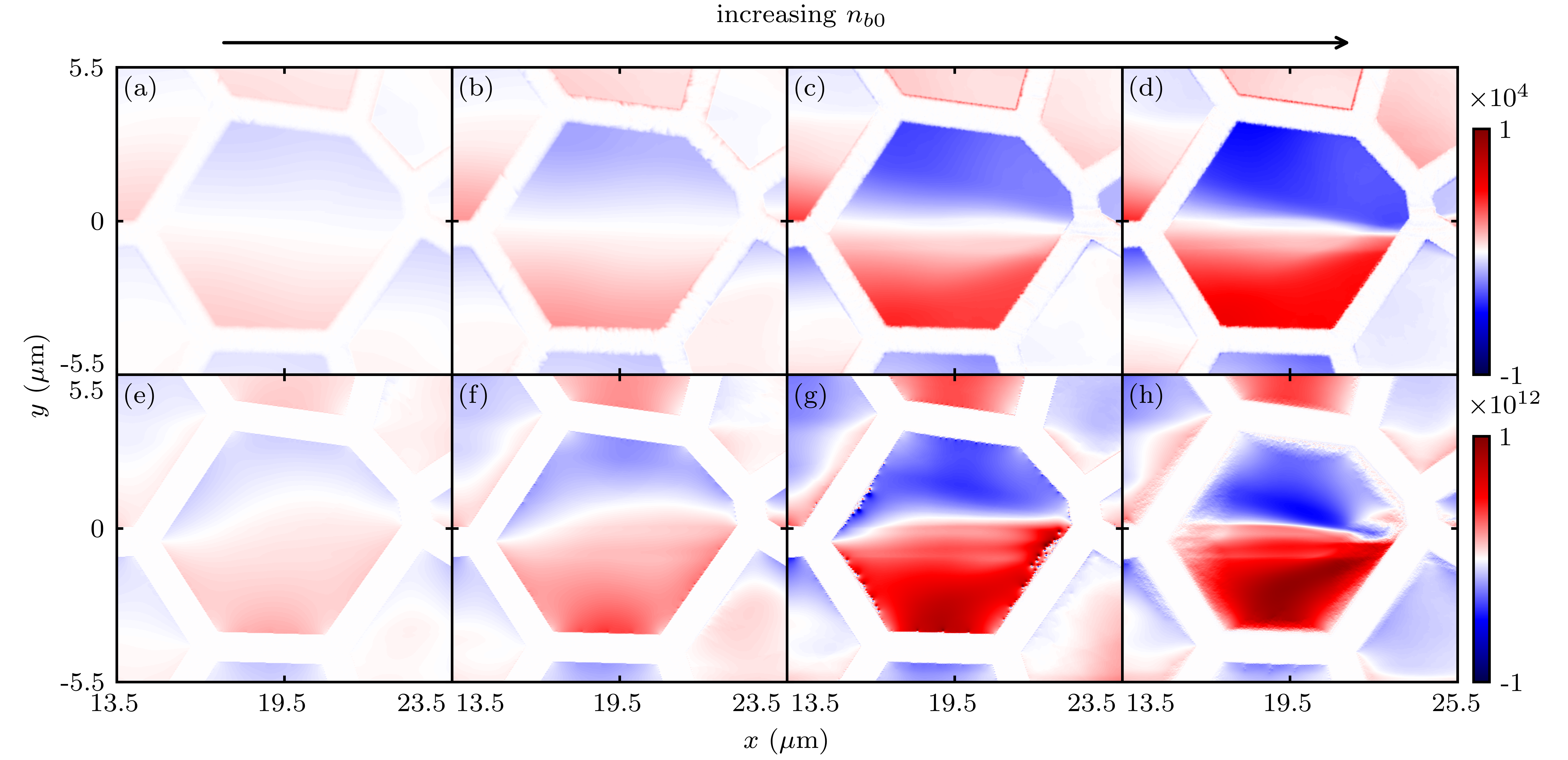} 
	\caption{From top to bottom, distributions of azimuthal magnetic field $B_{z}$ (in tesla) and transverse electric field $E_{y}$ (in V/m) in a selected region. From left to right, the initial density of the electron beam $n_{b0}=3.44\times10^{24}$, $6.88\times10^{24}$, $1.376\times10^{24}$, and $1.72\times10^{25}$ m$^{-3}$. In all the cases, $l_{c}$ keeps at $8\ \mathrm{\mu m}$.}
	\label{fig_2}
\end{figure*}

Recently, we reported that high-current relativistic electron beam (REB) propagation in porous materials (e.g., foam) can cascade into thin and dense branches at a length scale $d_0$ given by
\begin{eqnarray}
d_{0} \propto l_{c}^{-1/3}n_{b0}^{-2/3}\gamma^{2/3},\label{eq:eq1}
\end{eqnarray}
thereby bringing the branched flow phenomenon to high-energy-density physics \cite{Jiang1}. Here $l_{c}$ is the average pore size of the foam, $n_{b0}$ and $\gamma$ denote the initial density and Lorentz factor of the REB, respectively. Notably, REB branching differs from beam-plasma instabilities, as it arises from the microscopic skeleton-and-pore heterogeneity of the foam. The compensating return current, located within skin layers of the skeletons, induces magnetic fields in the unevenly distributed pores. As a consequence, the beam electrons undergo successive scattering by these fields, leading to branched flow. Furthermore, REB branching is accompanied by a significant increase in beam-target energy coupling efficiency, which can be two orders of magnitude larger than that in homogeneous plasmas with similar average density. 

In this paper, we present a comprehensive investigation into the branched flow of REBs, focusing on two aspects: (1) the impact of beam and foam parameters on the branching patterns, and (2) the dynamics of individual beam electrons. The paper is organized as follows: Section II introduces pore-resolved particle-in-cell (PIC) simulations, providing detailed results on the branched flow pattern. Since the branched flow pattern is determined by self-generated fields, Sect. III delves into the discussion on how REB and foam parameters affect these fields. Section IV presents the influence of interaction parameters on the branched flow pattern. In Sect. V, we present detailed analyses of the dynamics of individual beam electrons. The paper concludes with discussions in Sect. VI.

\section{Pore-Resolved Particle-in-Cell Simulation and Results}
In early simulations regarding REB transport in foam, the foam is usually simplified as a homogeneous continuum with a volumetric average density \cite{Jung,Fuchs,Manclossi,Romagnani,Ruyer}. That is, the foam's possible fine structuring is neglected. A likely reason for the assumption is that the foam considered is rather dense, such that its internal structure is smaller than the spatial scale of interest. Another reason might be the computational limitations at that time, which may have hindered the resolution of such fine structures. While results from the continuum assumption agree reasonably well with experiments using dense foam, this assumption proves inadequate for low-density foams characterized by spongelike submicron-sized intertwining solid-density skeletons with micron-sized empty pores in between \cite{Nagai}. These microstructures can significantly affect macroscopic results, necessitating their consideration in simulations \cite{YTLi,Robinson1,Chatterjee,Belyaev,HuangPPCF,Kemp,Cipriani,Jiang2}. 

To elucidate the impact of these microstructures on the propagation of REBs, we perform pore-resolved two-dimensional (2D) PIC simulations using the \textsc{epoch} code \cite{Arber}. As shown in Fig. \ref{fig_1}, in the simulation, the porous material is SiO$_{2}$ foam. The foam's internal structures are modeled by random Voronoi cells \cite{Song}. For definitiveness, the skeleton thickness $l_{d}=1\ \mathrm{\mu m}$ and average pore size $l_{c}=8\ \mathrm{\mu}$m. The skeletons consist of solid-density Si and O atoms, with respective densities of $n_{\mathrm{Si}}=2.15\times10^{28}$ m$^{-3}$ and $n_{\mathrm{O}}=4.3\times10^{28}$ m$^{-3}$. For simplicity, the initial REB is monoenergetic and of uniform density $n_{b0}=1.72\times10^{25}$ m$^{-3}$, with a duration of 400 fs and Lorentz factor $\gamma=100$. Detailed discussions regarding the impact of foam and REB parameters will follow. The simulation box spans from 0 $\mu$m $<x<$ 100 $\mu$m and -15 $\mu$m $<y<$ 15 $\mu$m. To resolve the skin depth of the solid skeletons, the spatial resolution $\Delta=12.5$ nm. There are 14 macroparticles per cell for both Si and O atoms and 100 for the beam electrons. In addition, over 300 macroelectrons per cell can be produced from the skeletons by field ionization. Periodic boundaries are used in the transverse direction. As discussed in Ref. [\onlinecite{Jiang1}], collisions are deemed negligible and are not included here.

\begin{figure}[!htb]
	\centering 
	\includegraphics[width=8.6cm]{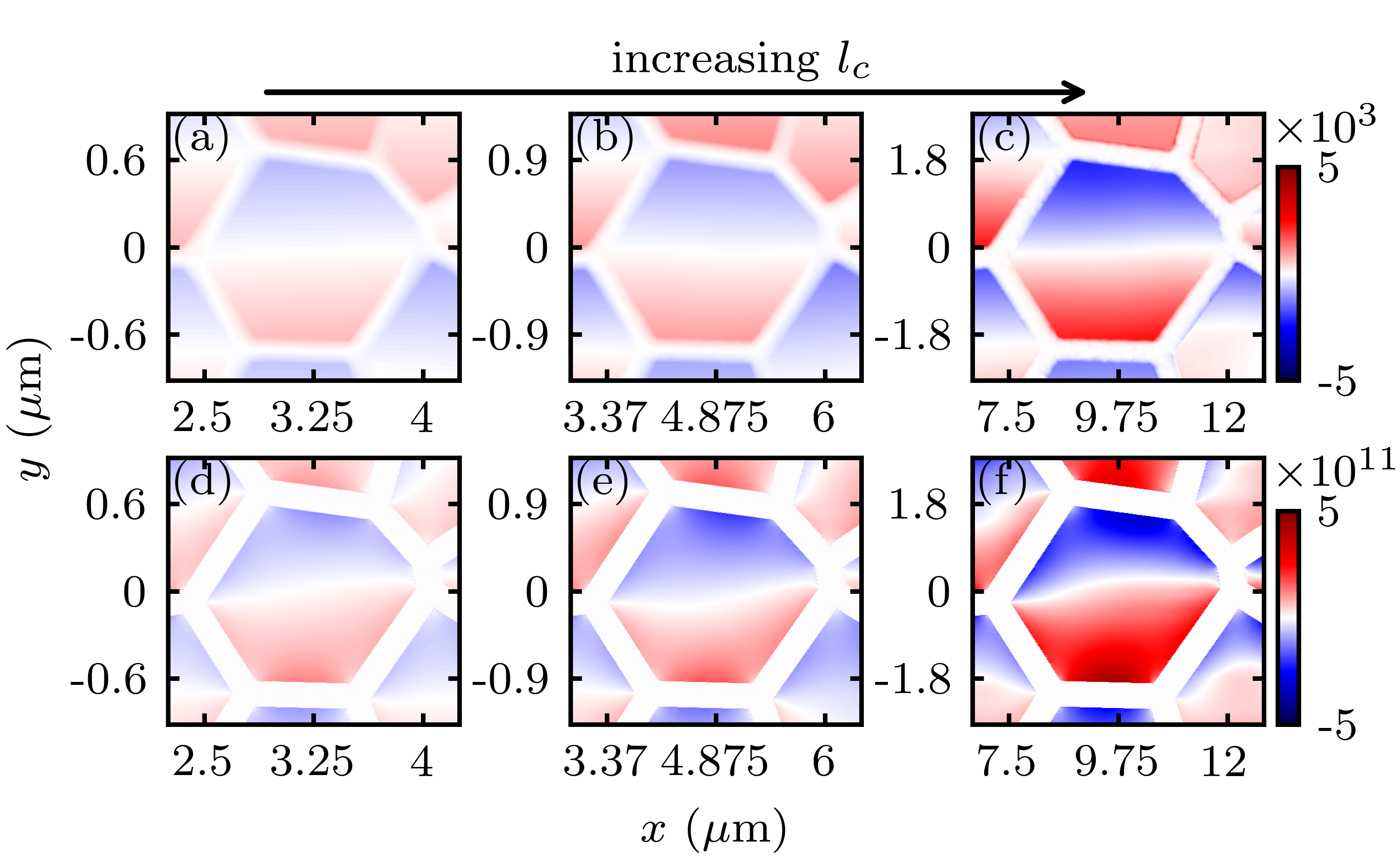} 
	\caption{From top to bottom, distributions of azimuthal magnetic field $B_{z}$ (in tesla) and transverse electric field $E_{y}$ (in V/m). From left to right, the average pore size of the foam $l_{c}=1.33$, $2$, $4$, and $8\ \mathrm{\mu m}$.}
	\label{fig_3}
\end{figure}

As shown in Fig. \ref{fig_1}, REB propagation in the foam induces uneven fields in the pores. The latter, in turn, deflects the beam electrons, resulting in REB branching. Specifically, the REB breaks up into three narrow dense branches. These branches subsequently broaden, intersect, and evolve into a fluctuating pattern. Since the Lorentz force associated with the azimuthal magnetic field $B_{z}$ is twice as large as that from electric fields (see discussion later and Ref. [\onlinecite{Jiang1}]), $B_{z}$ is more influential in shaping the dynamics of beam electrons. To visualize the electron motion, one may neglect the effects of electric fields and focus on the deflection caused by $B_{z}$ for simplicity. The gyroradius of beam electrons, given by $r_{g}=\gamma m_{e}c/eB_{z}\sim41.6$ $\mathrm{\mu}$m, is significantly larger than the average pore size (or the characteristic length scale of $B_{z}$) of $l_{c}=8$ $\mathrm{\mu}$m. Here $m_{e}$ and $-e$ are the electron rest mass and charge, and $c$ is the vacuum light speed. Therefore, the electrons can travel over several pores before significant deflection occurs. 

Considering the REB dynamics from an energy perspective may provide additional insights. The effective potential strength of the uneven fields is expressed as $V_{\mathrm{rms}}\sim\mu_{0}e^{2}c^{2}l_{c}^{2}n_{b0}/8\sqrt{3}$ \cite{Jiang1}, and the initial kinetic energy of the beam electron reads $E_{k,b0}=\gamma m_{e}c^{2}$. Here, $\mu_{0}$ is the vacuum permeability. The ratio $V_{\mathrm{rms}}/E_{k,b0}\sim 0.03\ll 1$, indicating that the magnitude of the effective potential is sufficiently small to cause only forward scatterings of the REB. These weak scatterings over several peaks and valleys of the potential $V$ lead to the formation of branched flows. This process is characterized by seemingly random, time-irreversible behavior, yet it exhibits deterministic and predictable tendencies. Since the distribution of pores, along with the associated effective potential, is isotropic, the transverse separations between two adjacent caustics are almost the same and match well with $l_{c}$. Therefore, branched flow of REBs presents a new regime of beam-plasma interaction with typical features of complex systems, where patterns emerge from random yet correlated interactions. 

\section{Self-Generated Pore Fields}

\begin{figure}[!tb]
	\centering 
	\includegraphics[width=8.6cm]{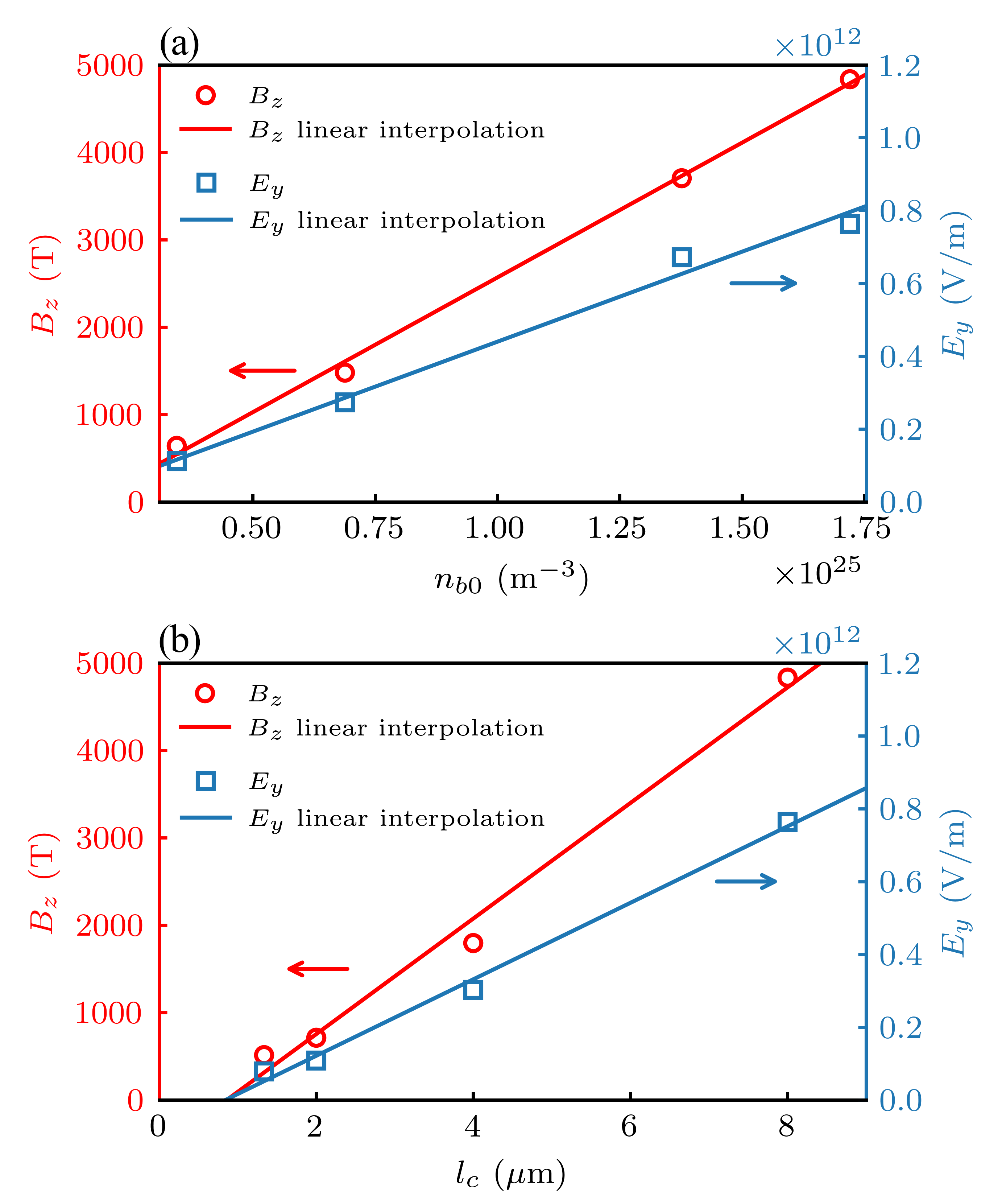} 
	\caption{(a) Strengths of the azimuthal magnetic field $B_{z}$ (red circles) and transverse electric field $E_{y}$ (blue squares) at $x=20\ \mathrm{\mu m}$ and $y=-3\ \mathrm{\mu m}$ obtained from simulations for different initial beam density $n_{b0}$ with $l_{c}=8\ \mathrm{\mu m}$. (b) Same as (a), but at $x=20l_{c}/8$ and $y=-3l_{c}/8$ obtained from simulations with different $l_{c}$. The solid lines are linear interpolations based on the simulation results.}
	\label{fig_4}
\end{figure}

\begin{figure*}[!htb]
	\centering
	\includegraphics[width=17.2cm]{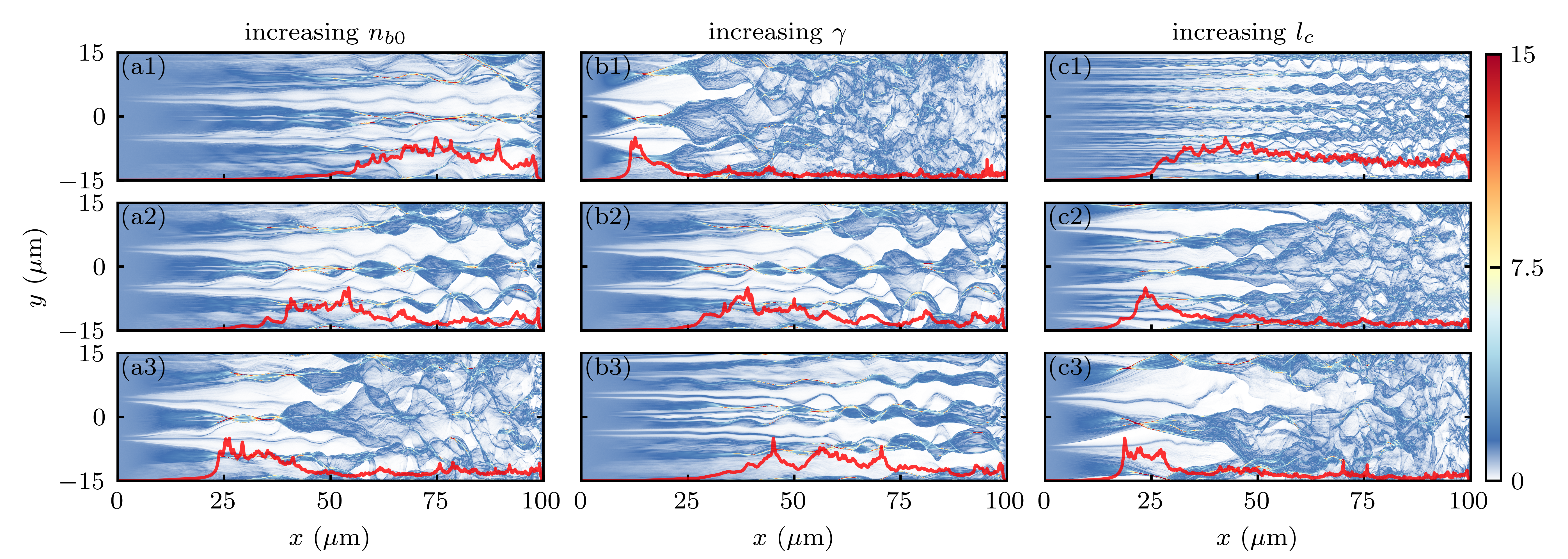} 
	\caption{Branched flow patterns (in unites of $n_{b0}$, same below) for (a1)-(a3) different initial beam densities at $n_{b0}=3.44\times10^{24}$, $6.88\times10^{24}$, and $1.38\times10^{25}$ m$^{-3}$ (with $\gamma=100$ and $l_{c}=8\ \mathrm{\mu m}$), (b1)-(b3) different beam Lorentz factors $\gamma=50$, 200, and 250 (with $n_{b0}=1.72\times10^{25}$ m$^{-3}$ and $l_{c}=8\ \mathrm{\mu m}$), (c1)-(c3) different average pore sizes $l_{c}=4$, 6, and 10 $\mathrm{\mu m}$ (with $\gamma=100$ and $n_{b0}=1.72\times10^{25}$ m$^{-3}$). Scintillation indices $\Sigma=\left(\left\langle n_{b}^{2}\right\rangle /\left\langle n_{b}\right\rangle ^{2}\right)-1$ corresponding to the beam electron densities are presented by red curves. The SiO$_{2}$ foams in (a) and (b) are of the same initial structure.}
	\label{fig_6} 
\end{figure*}

\begin{figure}[!htb]
	\centering 
	\includegraphics[width=8.6cm]{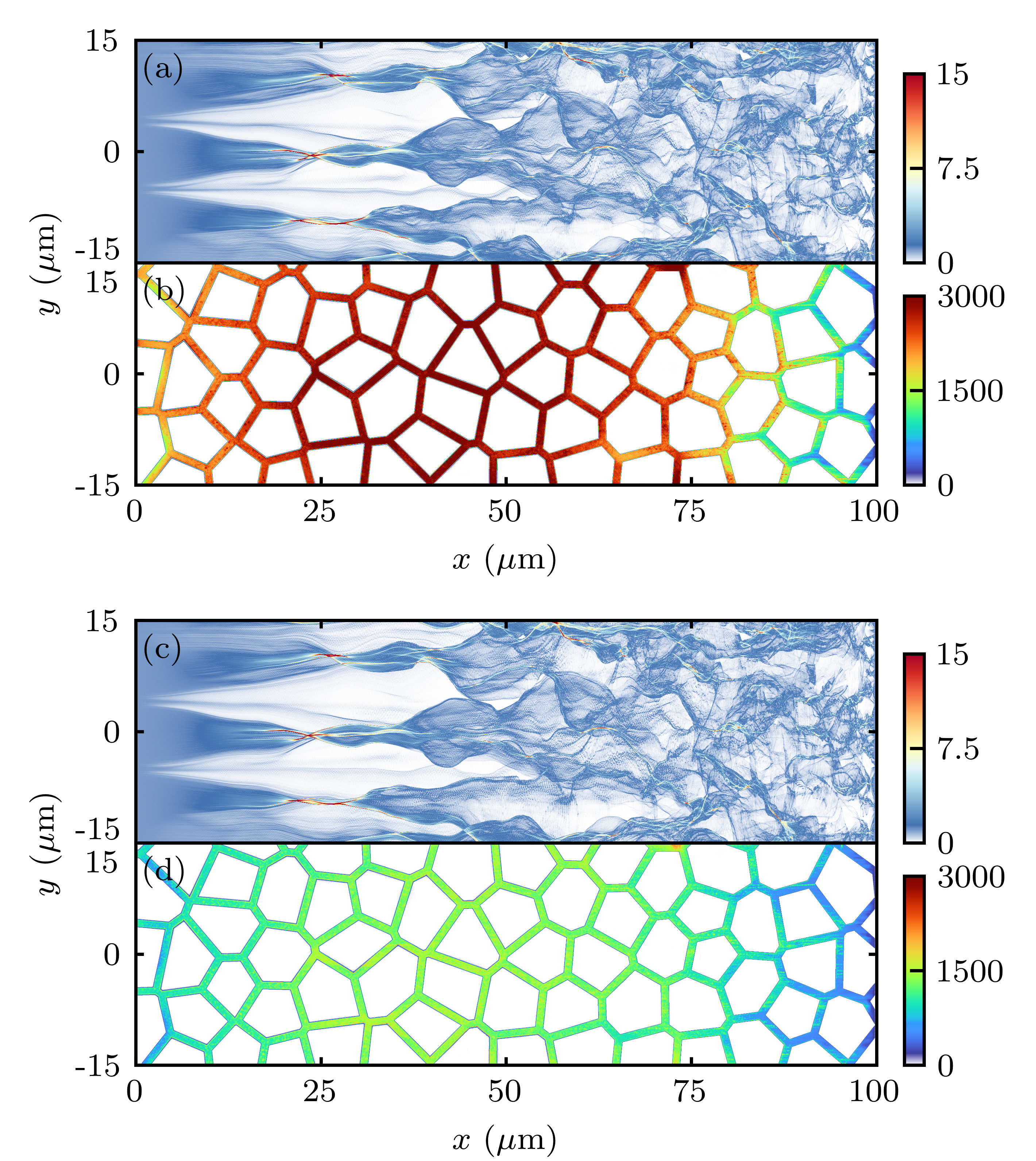} 
	\caption{(a)(c) Densities (in unites of $n_{b0}$, same below) of the beam electrons, (b)(d) densities of the foam electrons at $t=333$ fs. For (a)(b), the initial foam's Si atom density $n_{\mathrm{Si}}=4.3\times10^{27}$m$^{-3}$. For (c)(d), $n_{\mathrm{Si}}=2.15\times10^{27}$m$^{-3}$.}
	\label{fig_7}
\end{figure}

\begin{figure*}[!htb]
	\centering 
	\includegraphics[width=17.2cm]{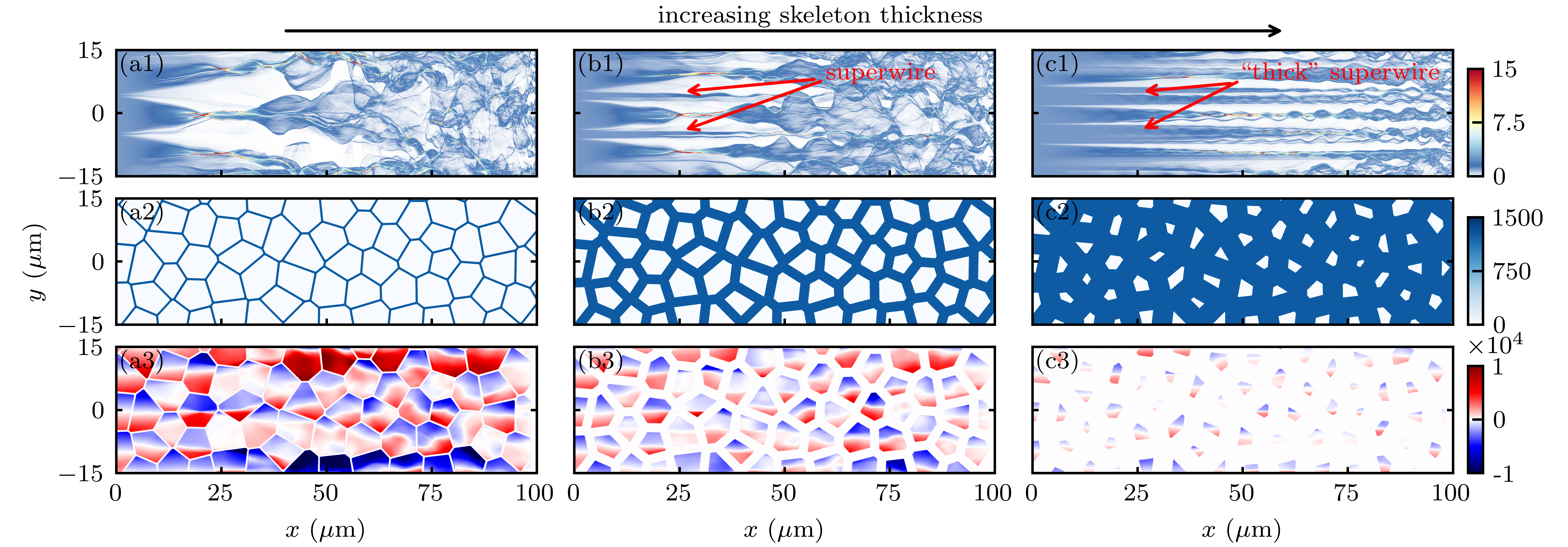} 
	\caption{From top to bottom, distributions of beam electron density (in unites of $n_{b0}$, same below), initial foam's Si atom density, and azimuthal magnetic field $B_{z}$ (in tesla). From left to right, the skeleton thickness $l_{d}=0.5$, 2.3, and 4.8 $\mathrm{\mu m}$.}
	\label{fig_8} 
\end{figure*}

Given that the self-generated pore fields $B_{z}$ and $E_{y}$ result in REB branching, this section considers the influence of interaction parameters on $B_{z}$ and $E_{y}$. As discussed in Ref. [\onlinecite{Jiang1}], $B_{z}$ arises from current neutralization. The return current, induced within skin layers of skeletons, are of huge density several times larger than the beam current, namely $j_{r}\sim|j_{b0}|l_{c}/2l_{s}\mathrm{exp}(1)$. Here $j_{b0}$ is the beam-current density, and  $l_{s}$ is the skin length of the skeletons. Since the solid-density skeletons effectively shield fields outside skin layers, the magnetic field induced by the current is distributed in the vacuum pores. The beam and return currents contribute approximately equally to the magnetic field, with the total magnetic strength given by Amp\'{e}re's law as $|B_{z}|\sim\mu_{0}en_{b0}l_{c}c/2\propto n_{b0}l_{c}$. In addition, the REB, being a bunch of electrons, provides the space charge for the electrostatic fields $E_{y}$ in the pore regions. From \mbox{Poisson' s} equation one obtains $E_{y}\sim en_{b0}l_{c}/4\varepsilon_{0}\propto n_{b0}l_{c}$. As both $B_{z}$ and $E_{y}$ originate from the REB itself, the time scale for their onset is $\sim l_{c}/c$ (several femtoseconds), much shorter than the expansion time scale of the skeletons. The magnetic field $B_{z}$ plays a more crucial role in scattering beam electrons ($|F_{y}|=e(c|B_{z}|-|E_{y}|)\sim ec|B_{z}|/2$), as the electrostatic and magnetic forces associated with the beam tend to cancel each other, leaving the net magnetic force provided by the return current to scatter the beam electrons.

The above analysis demonstrates a linear scaling between both $B_{z}$ and $E_{y}$ with $n_{b0}$ and $l_{c}$. This correlation is substantiated through a series of simulations with varied values of $n_{b0}$ and $l_{c}$. Figures \ref{fig_2} and \ref{fig_3} illustrate the observed increase in $B_{z}$ and $E_{y}$ with increasing $n_{b0}$ and $l_c$. In addition, in all the cases, the distributions of pore fields are quite similar. This further validates the association of pore fields with the skeleton-and-pore microstructure, indicating that the free evolution of these fields is constrained by the latter. Figure \ref{fig_4} for the field strengths at the same or equivalent positions shows the linear scaling of $B_{z}$ and $E_{y}$ with $n_{b0}$ and $l_{c}$, agreeing with our analysis.

\section{Branched Flow Patterns For Different Interaction Parameters}

\subsection{Influence of $n_{b0}$, $\gamma$, and $l_{c}$ on the branching pattern}

Since the pore fields and the associated effective potential strength $V_{\mathrm{rms}}$ depend on $n_{b0}$ and $l_{c}$, the resulting branched flow patterns are affected by these interaction parameters. In addition, the Lorentz factor $\gamma$ of the REB determines the dynamics of beam electrons under the pores fields, it is thus also crucial for the branched flow patterns. One can characterize the branched flow patterns by considering the distance $d_{0}$ from the foam's front surface to the first caustics, thus $d_0$, in a sense, represents the transition from order to randomness. Figure \ref{fig_6} shows the branched flow patterns and associated variation strength of the beam density (namely, scintillation index) $\Sigma=\left(\left\langle n_{b}^{2}\right\rangle /\left\langle n_{b}\right\rangle ^{2}\right)-1$ for different interaction parameters. It shows that $d_{0}$ decreases with increasing $n_{b0}$ and $l_{c}$ but increases with higher $\gamma$, in good agreement with Eq. \eqref{eq:eq1}. 

As $d_{0}$ increases, we find that the branches are less likely to overlap with each other. Instead, they seem to be confined in the pores and propagate rather stably as individual strands, leading to a mild spatial evolution of $\Sigma$. This behavior can be interpreted in terms of the focusing angle $\theta$ of the REB, defined as $\tan\theta=l_{c}/2d_{0}\propto n_{b0}^{2/3}l_{c}^{4/3}\gamma^{-2/3}$. We see the branches become collimated with decreasing $n_{b0}$ and $l_{c}$, and increasing $\gamma$, which agrees with Fig. \ref{fig_6}. Collimated branches may lead to low-divergence $x$- and $\gamma$-rays.

It is worth mentioning that $l_{c}$ not only affects $d_{0}$ and $\theta$, but also determines the number of the first caustics in the transverse direction. It is consistent with the universal feature of branched flows, where the average transverse separation between two adjacent caustics equals the correlation length of the effective potential \cite{Heller,Kaplan,Metzger,Metzger1} - corresponding, in our case, to $l_{c}$. As $l_{c}$ approaches zero, the transverse separation between adjacent caustics tends to zero, and $d_{0}$ approaches infinity. That is, REB branching is absent in continuum media.

\subsection{Influence of $n_{p}$ and $l_d$ on the branching pattern}

Equation \eqref{eq:eq1} reveals another unique feature of REB branching: the branched flow pattern is independent of the skeleton electron density $n_{p}$. This distinguishes REB branching from beam-plasma instabilities, as the latter can be sensitive to $n_p$ \cite{Weibel,Watson,Davidson,Gremillet,Gremillet1,Krasheninnikov,Bret,Bret1}. This feature of REB branching is observed under the condition that the skeleton is sufficiently dense to be considered rigid during REB propagation, i.e., $n_{p}\gg n_{b0}$. In this case, the pore fields and associated $V_{\mathrm{rms}}$ exclusively depend on $n_{b0}$ and $l_{c}$, as discussed in Sect. III. Therefore, the dynamics of beam electrons are unaffected by $n_{p}$. Figure \ref{fig_7} for simulation results with different skeleton atom densities confirms that the branching patterns remain almost identical despite substantial differences in the skeleton densities.

Figure \ref{fig_8} illustrates the influence of skeleton thickness $l_{d}$ on the branching pattern. This effect stems from our model, where $V_{\mathrm{rms}}$ is obtained under the assumption that $l_{d}$ approaches zero. Given that both $B_{z}$ and $E_{y}$ are effectively zero in the skeletons, an increase in $l_{d}$ results in decreasing $V_{\mathrm{rms}}$. As a result, $d_{0}$ occurs at a longer distance in dense foam with thicker skeletons. In addition, finite $l_{d}$ gives rise to the formation of so-called superwires \cite{Daza}, which can propagate as stable thin filaments over extended distances, as indicated by the red arrows in Figs. \ref{fig_8}(b1) and (c1). Superwires emerge when a certain beam electron and its neighboring ones in phase space coincidentally follow similar trajectories, maintaining stable manifolds over a finite time. As $\partial_{xx} V$ and $\partial_{yy} V$ are effectively zero in the skeletons, the stability matrix $\mathcal{M}$ of beam electrons with similar conditions is more likely to remain stable for thick skeletons. See Ref. [\onlinecite{Mattheakis}] for the definition of $\mathcal{M}$ and detailed discussion. That is, superwires can form and propagate extensively in foams with thick skeletons.

\section{From Individual Beam Electrons' Perspective}

\begin{figure}[!htb]
	\centering 
	\includegraphics[width=8.6cm]{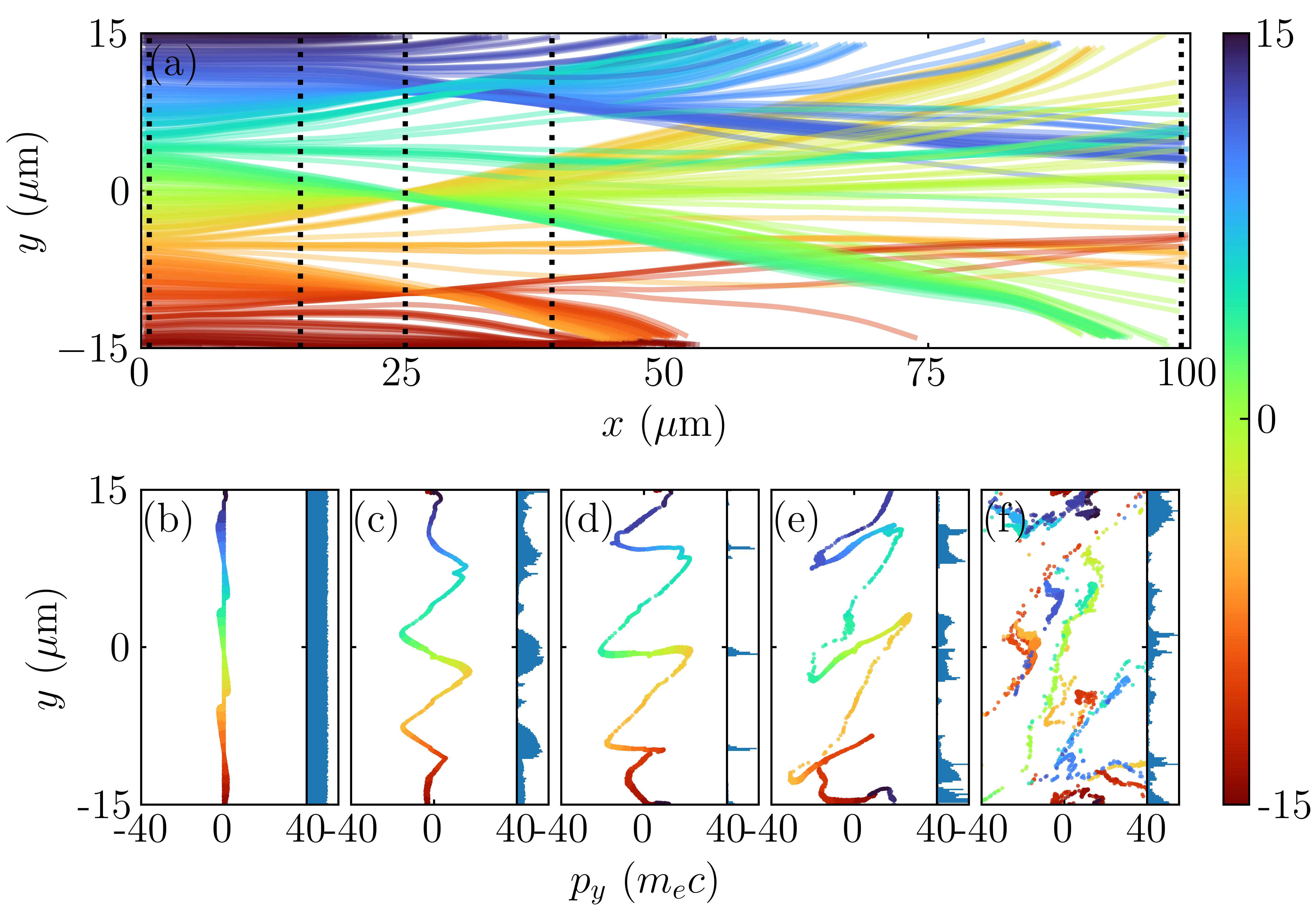} 
	\caption{(a) Trajectories of randomly selected 300 beam electrons. (b)-(f) Phase spaces ($p_{y},y$) of randomly selected 4000 beam electrons at $x=0.8$, 15.2, 25.2, 39.2, and 99.2 $\mathrm{\mu m}$, respectively. The corresponding histograms are shown on the right side. The color bar is for the inital $y$ coordinates of the selected beam electrons.} 
	\label{fig_9} 
\end{figure}

In this section, we employ particle tracing methods to gain insights into REB branching. As shown in Fig. \ref{fig_9}(a), the trajectories of individual electrons exhibit a quasi-ballistic nature. That is, the branched flow pattern develops linearly, allowing us to treat the beam electrons as single particles due to the absence of collective motion. Here, the term ``linearly" refers to that the uneven background fields remain quasistatic during pattern formation. In this sense, one may artificially divide the REB branching into two stages. The first is a nonlinear stage of the background fields induction by the return current, as a consequence of the collective response of foam electrons to the high-current REB. The second stage appears to be linear, during which the beam electrons undergo scattering by the uneven fields, behaving as individual particles.

The formation of branching patterns from random scatterings is elucidating when observed in the phase space $(p_{y},y)$ of the traced beam electrons. As shown in Fig. \ref{fig_9}(b), the initial $(p_{y},y)$ is nearly a straight line since the beam is initially monoenergetic. The corresponding beam density is also uniform, as indicated by the histogram on the right side. As the beam propagates, successive random kicks from the uneven fields cause the phase space $(p_{y},y)$ to stretch and curve, resulting in a focused beam, see Fig. \ref{fig_9}(c). As a result, three singularities, where $\partial_y p_{y}=\infty$, occur around $x\sim25\ \mathrm{\mu m}$, as shown in Fig. \ref{fig_9}(d). These singularities in the phase space correspond to caustics, where the beam converges to an extremely high density, as shown in the histogram and Fig. \ref{fig_1}. The phase space undergoes folding after the caustics, giving rise to individual branches in the coordinate space. For each branch, the peripheral density is higher than that at the center, see Fig. \ref{fig_9}(e). Eventually, the phase space evolves into random distributions through further random scatterings as well as overlapping of different branches, see Fig. \ref{fig_9}(f). Figure \ref{fig_9}(f) also reveals an intriguing aspect of REB branching wherein groups (or manifolds) of beam electrons at neighboring initial locations can extend over significant zones over time. This emphasizes the sensitivity of a single electron's motion to its initial condition. However, the resultant branched flow pattern, formed by the ensemble of beam electrons, adheres to well-defined statistical laws \cite{Kaplan,Metzger,Metzger1}.

\begin{figure}[!htb]
	\centering 
	\includegraphics[width=8.6cm]{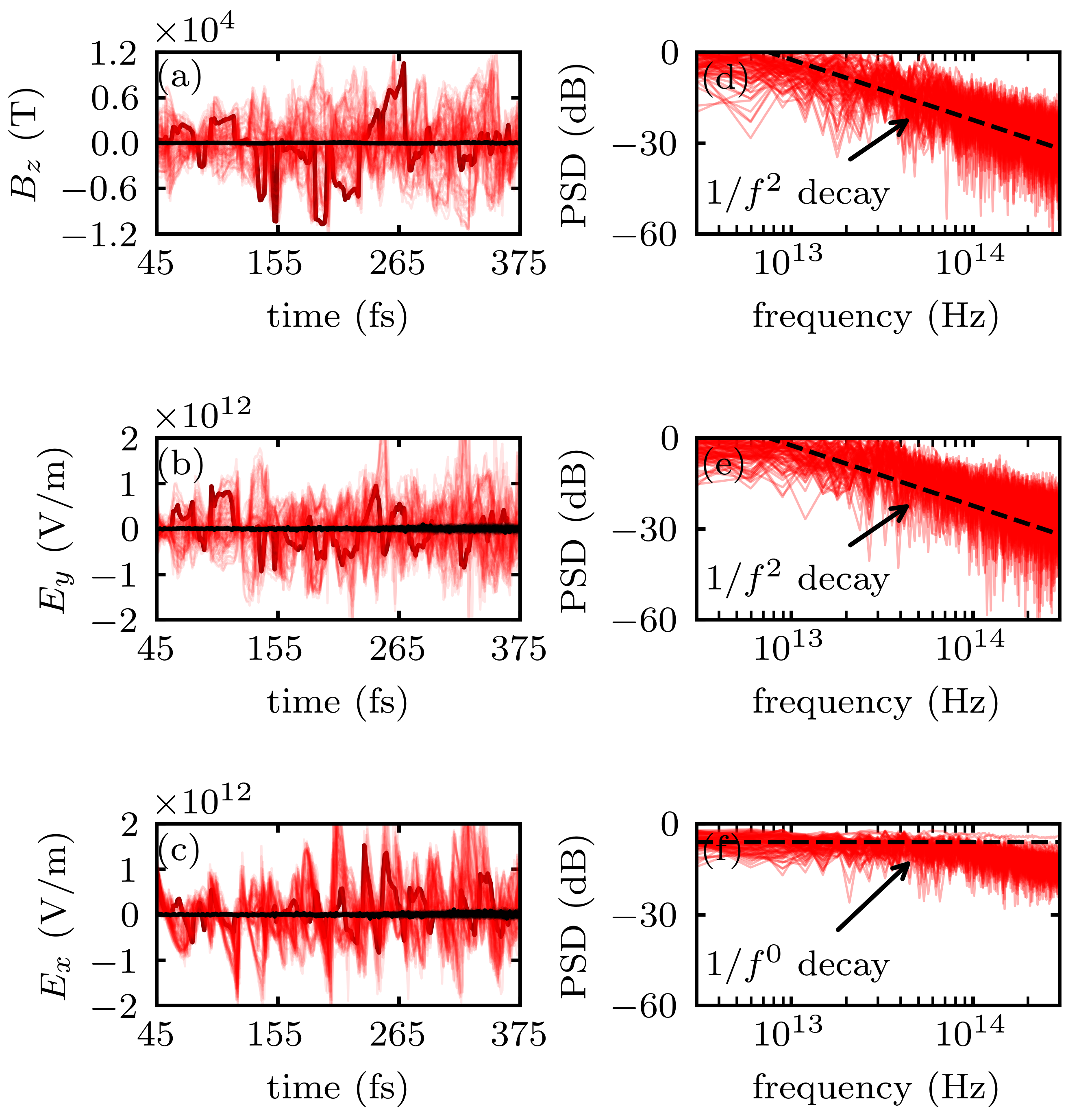} 
	\caption{Evolution of (a) azimuthal magnetic field $B_{z}$, (b) transverse, and (c) longitudinal electric field $E_{y}$ and $E_{x}$ experienced by randomly selected 100 beam electrons as they propagate in the foam (red) and a homogeneous target (black) at the same average density of the foam for reference. The dark red curves correspond to that experienced by a typical beam electron. (d)-(f) show the corresponding power spectral densities (PSD) of $B_{z}$, $E_{y}$, and $E_{x}$ in the case of foam target, respectively.}
	\label{fig_10}
\end{figure}

Figures \ref{fig_10}(a)-(c) show the fields experienced by beam electrons. Indeed, in the foam target, these fields evolve randomly and are several orders of magnitude stronger than that of homogeneous targets at the same average density. Such strong and dynamic fields may lead to efficient radiation in foam targets. By analyzing the power spectral density, defined as $\mathrm{PSD}=\mathcal{F}(\mathcal{W})^{2}/C$, one sees in Figs. \ref{fig_10}(d) and (e) that the frequency densities of $B_{z}$ and $E_{y}$ are inversely proportional to $f^{2}$. It further confirms the time-correlated characteristic, akin to Brownian noise \cite{Ohtomo}, of the transverse fields. Here. $\mathcal{F}$ denotes to Fourier transform, $\mathcal{W}$ is a generalized signal corresponding to the fields experienced by beam electrons in this context, $C$ serves as a normalizing constant, and $f$ is the frequency of $\mathcal{W}$. On the other hand, the frequency density of $E_{x}$ appears relatively constant, namely inversely proportional to $f^{0}$, as shown in  Figs. \ref{fig_10}(f). This suggests a negligible correlation in the longitudinal direction, resembling white noise characteristics \cite{Ohtomo}. Therefore, even though the beam electrons experience random interactions in both longitudinal and transverse directions, branching exclusively manifests in the latter.

\section{Discussion}

REB branching in porous materials is a quite complicated process. This paper only addresses REBs with simplified ``plane wave'' profile, focusing on the impact of beam and foam parameters on the branched flow pattern. There are other important interaction parameters. For example, REB's temporal profile is directly related to the time-dependent $n_{b0}(t)$. Therefore, the location of the first caustics can vary over time as $d_{0}\propto l_{c}^{-1/3}n_{b0}(t)^{-2/3}\gamma^{2/3}$. In the case of a temporally Gaussian REB with sufficient duration, $d_{0}$ first decreases with time and then increases with it. The initial emittance and energy spread of REBs can also affect the resulting pattern \cite{Jiang1}. Moreover, foams are not necessarily isotropic, i.e., $l_{c}$ can vary in different locations, modifying the correlation of induced pore fields and resulting REB dynamics. 

Branched flow of REBs with unique features may open up new avenues in beam-plasma physics. Due to the pore-and-skeleton structures, the foam facilitates effective and volumetric heating \cite{Jiang1}, thereby attaining high-energy-density states. The heating mechanism for ions may depend on the skeleton thickness, probably transitioning from skeleton-normal sheath acceleration for thick skeletons to Coloumb explosion for thin skeletons. Moreover, there is a prospect of fabricating the foam target using nuclides of interest and harnessing the dense return current within the skeletons for nuclear excitation. In addition, REB propagation in foam experiences rapidly changing strong fields, offering promise for generating tunable bright radiations by adjusting beam and foam parameters. We believe that the journey to understanding this intriguing phenomenon has just begun.

\section*{Acknowlegdement}

This work is supported by the National Key R\&D Program of China (Grant No. 2022YFA1603300), the National Natural Science Foundation of China (Grants No. 12205201, No. 12175154, No. 12005149, and No. 11975214), the Shenzhen Science and Technology Program (Grant No. RCYX20221008092851073, 20231127181320001). The \textsc{epoch} code is used under
UK EPSRC contract (No. EP/G055165/1 and No. EP/G056803/1).

\section*{Data Availability}
The data that support the findings of this study are available from the corresponding author upon reasonable request.

\end{document}